\documentclass[11pt]{article}
\usepackage{geometry}     % See geometry.pdf to learn the layout options. There are lots.
\usepackage{graphicx}
\usepackage{amsmath,amssymb}  % Better maths support & more symbols
\usepackage{bm}  % Define \bm{} to use bold math fonts
\usepackage{slashed}
\newcommand\beq{\begin{equation}}
\newcommand\eeq{\end{equation}}
\newcommand\Tr{\text{Tr}}

\newcommand\vev[1]{\langle #1\rangle}
\newcommand\diag{\text{diag}}
\newcommand\hc{\text{h.c.}}
\newcommand\LLH{$\text{L}^2\text{H}$}
\title{Top quark induced vacuum misalignment in little higgs models}
\author{Benjam\'\i{}n Grinstein\thanks{bgrinstein@ucsd.edu} \, and Michael Trott\thanks{mrtrott@physics.ucsd.edu}\\
University of California, San Diego; \\ Dept. of Physics, La Jolla, CA 92093-0315}
\date{\today \vskip-2in \hfill UCSD PTH 08-07\vskip2in}            % Activate to display a given date or no date

\begin{document}
\maketitle
\begin{abstract}
We show that  the effect of the top quark can
dominate over the effect of the gauge sector in determining the vacuum alignment in little higgs (LH) models.
We demonstrate that in the littlest LH model and the $SU(2) \times SU(2) \times U(1)$ LH model,
ensuring that the correct vacuum alignment is chosen requires that a subset of the gauge sector couplings be large to overcome the effect of the top quark. We 
quantify this effect by deriving bounds on the couplings in the gauge sector and demonstrate that these bounds provide a compelling theoretical reason for the gauge coupling constant hierarchy in the $SU(2) \times SU(2) \times U(1)$ model that reduces the Goldstone decay constant scale to a $\rm TeV$. We also argue that for a class of LH models with T parity the top quark drives the correct vacuum alignment and therefore all gauge couplings can be small.

\end{abstract}
\newpage
\section{Introduction}
%\subsection{}

Little Higgs (LH) models offer an alternative to the standard model in
which no fundamental scalars need be introduced (for reviews see
\cite{Schmaltz:2005ky}).  Generally, in LH models the Higgs is a composite
particle, bound by interactions that become strong at a scale
$\Lambda$. The mass of the Higgs is much less than $\Lambda$ as
the Higgs is a pseudo-Goldstone boson (PGB) of broken
global symmetries in the theory of the new strong interaction.

The global ``flavor'' symmetry $G_f$ of these models has a subgroup
$G_w$ that is weakly gauged. In the absence of this weak gauge force,
the flavor symmetry is broken spontaneously to a subgroup $H$ due to the strong interactions at the scale $\Lambda$. As a result,
there are massless Goldstone bosons that are coordinates on the
$G_f/H$ coset space. Including the effect of $G_w$ as a perturbation,
a particular vacuum can be selected in $G_f/H$ which has a particular
calculable spectrum of Goldstone bosons.  Determining the vacuum
selected due to the $G_w$ perturbation is the ``Vacuum Alignment''
problem \cite{Preskill:1980mz}.

Whether part or all of $G_w$ is spontaneously broken depends on the
vacuum alignment.  Once the vacuum degeneracy is lifted by the $G_w$
perturbation, there are no more massless Goldstone Bosons; instead
there are would-be Goldstone bosons that are eaten by the broken
generators of $G_w$ and PGB's whose masses vanish with the couplings
in $G_w$. The Higgs is  the lightest PGB in LH models, and
its mass is naturally much less than $\Lambda$ due to the
collective symmetry breaking mechanism.

Realistic models identify $G_w$ with the electroweak interactions, or
with a larger group that contains the electroweak interactions. They
also include interactions responsible for quark masses. It is
customary to address the vacuum alignment problem in LH models by
first analysing the effect of $G_w$ and then verifying that the effect
of other interactions, like those giving quarks their masses, do
not destabilize the solution.

However the top quark Yukawa coupling is larger than the
electroweak gauge couplings. The top quark effect on vacuum alignment
can be larger than that of the weakly gauged interactions, even dominant. If in the
absence of $G_w$ the selected vacuum is different from the one chosen
by $G_w$ alone, and if this different vacuum leads to the wrong low
energy spectrum, then the only way to insure the vacuum that gives the SM at low energies is
selected is to make the gauge couplings in $G_w$ large enough that
they dominate the top quark effect.

In this paper we explore this observation explicitly in the littlest
higgs\cite{ArkaniHamed:2002qy} model ($\text{L}^2\text{H}$) and two of
its variants.  In the \LLH\ we will show that to lowest order in the
top Yukawa there are two inequivalent degenerate vacua, a ``good''
vacuum alignment ($\Sigma_{ew}$) that contains the SM electroweak theory in its
low energy limit and a ``bad'' vacuum alignment ($\Sigma_{B}$) that does not. We
then show that to next order in the top Yukawa, the ``bad'' vacuum is
favoured. Including the effect of gauge interactions, we derive a
bound on the gauge couplings that must be satisfied if the model is to
have $\Sigma_{ew}$ as a global minimum.  Having established the need
to consider the effects of the top on vacuum alignment in the 
\LLH\ model, we then consider the effect of the top on vacuum
misalignment in variants of the \LLH\ model that are more
phenomenologically viable.  We derive another bound for the $SU(2)
\times SU(2) \times U(1)$ LH model \cite{Perelstein:2003wd} and note
that the bound supplies a reason for the gauge coupling hierarchy $g_{1} \gg g_{2}$
that is desirable as it minimizes the fine tuning of the Higgs mass. We also argue that for
LH models with $T$-parity, the $\Sigma_{ew}$
vacuum is the absolute minimum at lowest order in the top-Yukawa,
demonstrating that for some LH model variants, the vacuum selected by
the gauge sector can be valid for small gauge couplings.

\section{Top vacuum misalignment for \LLH}
To establish notation we briefly review elements of the \LLH \,  \cite{ArkaniHamed:2002qy}. It has
$G_f=SU(5)$, $H=SO(5)$ and $G_w=\prod_{i=1,2}SU(2)_i\times U(1)_i$.
Symmetry breaking $SU(5)\to SO(5)$ is characterized by the Goldstone
boson decay constant $F$. The embedding of $G_w$ in $G_f$ is fixed by
taking the generators of $SU(2)_1$ and $SU(2)_2$ to be
\begin{equation}
Q^a_1=\begin{pmatrix}\frac{1}{2}\tau^a&0_{2\times3}\\0_{3\times2}&0_{3\times3}\end{pmatrix}
\qquad\text{and}\qquad
Q^a_2=\begin{pmatrix}0_{3\times3}&0_{3\times2}\\0_{2\times3}&-\frac{1}{2}\tau^{a*}\end{pmatrix}
\end{equation}
and the generators of the $U(1)_1$ and $U(1)_2$ 
\begin{equation}
Y_1=\frac1{10}\diag(-3,-3,2,2,2)\qquad\text{and}\qquad
Y_2=\frac1{10}\diag(-2,-2,-2,3,3). 
\end{equation}

The vacuum manifold is characterized by a unitary, symmetric
$5\times5$ matrix $\Sigma$. We denote by $g_{i}$ ($g'_{i}$) the gauge couplings associated
with $SU(2)_i$ ($U(1)_i$). If one sets $g_{1}=g_1'=0$ the model has an
exact global $SU(3)$ symmetry (acting on upper $3 \times 3$ block of
$\Sigma$), while for $g_{2}=g_2'=0$ it has a different exact global $SU(3)$
symmetry (acting on the lower $3 \times 3$ block). This gives rise to the
collective symmetry that ensures the absence of 1-loop quadratic
divergences in the higgs mass. To lowest order in the $G_w$ couplings, the
quadratically divergent contribution to the vacuum energy is
\begin{equation}
\label{eq:max-iso-v}
V_w(\Sigma)=\frac34 \, c \, F^4  \, \sum_\alpha g_\alpha^2 \Tr\left(T^\alpha\Sigma (T^\alpha)^T\Sigma^\dagger\right),
\end{equation}
where the sum on $\alpha$ runs over all generators of $G_w$. We have
normalized so that $c=1$ corresponds to the quadratic divergence
in the Coleman-Weinberg potential with a Euclidean momentum cut-off
$\Lambda=4\pi F$. 

It is standard to introduce the top quark so that the
collective symmetry argument still applies. Additional spinor fields
are introduced:  $q_R$, $u_L$ and $u_R$ that
transform as ${\bf{1}}_{2/3}$ under $SU(2)_1\times U(1)_1$, and $q_L$
transforming as ${\bf{2}}_{1/6}$. These couple  via
\begin{equation}
\label{eq:Ltop}
{\cal L}_{\text{top}}=-\frac12 \, \lambda_1 \, F \, \bar  \chi_{Li}^{\phantom{\dagger}}\, \epsilon_{ijk} \, \epsilon_{mn} \, 
\Sigma_{jm} \, \Sigma_{kn} \ q_R^{\phantom{\dagger}} - \lambda_2 \, F \, \bar u_L^{\phantom{\dagger}}\, u_R^{\phantom{\dagger}} +\hc
\end{equation}
where the indexes $i,j,k$ run over 1,2,3, the indexes $m,n$ over $4,5$
and the triplet $\chi_L^{\phantom{\dagger}}$ is 
\begin{equation}
\chi_L^{\phantom{\dagger}}=\begin{pmatrix}-i\tau^2 q_L^{\phantom{\dagger}}\\u_L^{\phantom{\dagger}}\end{pmatrix}.
\end{equation}
The vacuum energy is determined by the Coleman-Weinberg potential. Using a
momentum cut-off in Euclidean space, $|p_E|\le\Lambda$, it is given by
\begin{equation}
\label{eq:Vpsi}
V_t(\Sigma)=-\frac{N_c}{16\pi^2}\left[2\Lambda^2\Tr M^\dagger M
  +\Tr(MM^\dagger)^2\ln(MM^\dagger/\Lambda^2)-\frac12\Tr(MM^\dagger)^2\right],
\end{equation}
where $M=M(\Sigma)$ is the spinor mass matrix from Eq.~\eqref{eq:Ltop}
and $N_c=3$ is the number of colors. The quadratic and logarithmic
divergences are cut-off by modes of the UV completion of the
model. Even if we specified the UV completion we would be unable to
compute the precise cut-offs, so we parametrize them using $\Lambda=4\pi
F$ and two unknown constants:
\begin{equation}
\label{eq:Vpsi2}
V_t(\Sigma)=-\frac{N_c}{16\pi^2}\left[2c'\Lambda^2\Tr M^\dagger M
  +\Tr(MM^\dagger)^2\ln(MM^\dagger/\Lambda^2)-\frac12\hat c'\Tr(MM^\dagger)^2\right].
\end{equation}

The vacuum energy $V_t$ has two vacua\footnote{Considering $G_w$ global transformations on the vacuum
  matrix representations one can show that these are the only two
  physically distinct degenerate vacua.} that are degenerate 
at leading order. The vacuum alignment that  leads to $\prod_{i=1,2}SU(2)_i\times U(1)_i\to SU(2)\times U(1) $ is 
\begin{equation}
\label{eq:Sigma0}
\Sigma_{ew}=\begin{pmatrix} 0&0& \mathbf{1}_{2\times2}\\ 0 & 1& 0\\ \mathbf{1}_{2\times2} &0 &0\end{pmatrix}.
\end{equation}
The second vacuum alignment
 that is degenerate with $\Sigma_{ew}$ at leading order  in $V_t$ is
\begin{equation}
\Sigma_{B} =\begin{pmatrix}0&0&0&0&1\\ 0&1&0&0&0\\ 0&0&0&1&0\\ 0&0&1&0&0\\
  1&0&0&0&0\end{pmatrix}.
\end{equation}

The $\Sigma_{B}$ vacuum alignment leads to  $\prod_{i=1,2}SU(2)_i\times
U(1)_i\to U(1)\times U(1) $.
The difference in the vacuum energy between $\Sigma_{ew}$ and  $\Sigma_{B}$ is
\begin{multline}
V_t(\Sigma_{ew})-V_t(\Sigma_{B})=-\frac{N_c}{16\pi^2}\left[
(\lambda_1^2+\lambda_2^2)^2F^4
\left(\ln\left(\frac{(\lambda_1^2+\lambda_2^2)F^2}{\Lambda^2}\right)-\frac{\hat
    c'}2\right)\right.
\\\left.
-\left(\lambda_1^4F^4\ln\left(\frac{\lambda_1^2F^2}{\Lambda^2}\right)
+\lambda_2^4F^4\ln\left(\frac{\lambda_2^2F^2}{\Lambda^2}\right)-\frac{\hat
  c'}2(\lambda_1^4+\lambda_2^4)F^4
\right)\right].
\end{multline}
That this is positive is most easily seen by considering $\Lambda\gg
\lambda_{1,2}F$:
\begin{equation}
V_t(\Sigma_{ew})-V_t(\Sigma_{B})\approx 
\frac{N_c}{8\pi^2} \, 
\lambda_1^2 \, \lambda_2^2 \, F^4 \, \left[\ln\left(\frac{\Lambda^2}{(\lambda_1^2+\lambda_2^2)F^2}\right)+\frac{\hat
    c'}2\right]>0.
\end{equation}
The difference could be negative, and the $\Sigma_{ew}$ vacuum deeper, if
$\hat c'$ were large and negative. However,  $\hat c'$ is expected to be
positive, since it corresponds to a shift in the cut-off.

The $\Sigma_{ew}$ vacuum can be restored through the effects of the weak gauge
interactions.  The vacuum energy from \eqref{eq:max-iso-v} gives an
additional contribution to the energy difference
\begin{equation}
V_{w}(\Sigma_{ew})-V_w(\Sigma_{B})=-\frac{3}{16} \, c \, F^4 \, \left[g_1^{\prime2}+g_1^2\right].
\end{equation}
Combining results we obtain  the condition for the $\Sigma_{ew}$
vacuum alignment to be deeper than the  $\Sigma_{B}^2$ alignment 
is (for  $\Lambda\gg \lambda_{1,2}F$) :
\begin{equation}
\label{eq:bound1}
g_1^{\prime2}+g_1^2 >\frac{2N_c}{3\pi^2c} \, \lambda_1^2 \, \lambda_2^2 \, 
\left[\ln\left(\frac{\Lambda^2}{(\lambda_1^2+\lambda_2^2)F^2}\right)+\frac{\hat
    c'}2\right].
\end{equation}
Note that the vacuum alignment bound only restricts the gauge couplings $g'_1,g_1$, as 
both of the vacua $\Sigma_{ew}$ and $\Sigma_{B}$ have ${\bf{0}}_{2 \times 2}$ in the the lower right 
block of the vacuum alignment matrix. Thus calculations of higher order corrections to the bound will 
still only restrict these two couplings. We will consider some physical implications of this fact in the following two sections.

\subsection{Phenomenological implications of top vacuum misalignment in \LLH}

The \LLH\  is phenomenologically disfavoured \cite{Han:2003wu}  by electroweak
precision data (EWPD) but is an excellent toy model
to examine some consequences of this new constraint for LH model building.
In the  $\Sigma_{ew}$ vacuum alignment the top quark Yukawa is given by
\begin{equation} \label{yukawa}
\lambda_t=\frac{\sqrt2\lambda_1\lambda_2}{\sqrt{\lambda_1^2+\lambda_2^2}} + \mathcal{O}(\frac{v^2}{F^2}). 
\end{equation}
Using $ \lambda_t=\sqrt2 \, m_t/v$, along with $\Lambda = 4 \, \pi F$  the bound is
\begin{equation}
\label{eq:bound1a}
g_1^{\prime2}+g_1^2>\frac{2N_c}{3\pi^2c} \, \lambda_1^2 \, \lambda_2^2 \, 
\left[2 \, \ln\left(\frac{4 \, \pi \, m_t}{\lambda_1 \, \lambda_2 \, v}\right)+\frac{\hat
    c'}2\right].
\end{equation}
Minimizing \eqref{eq:bound1a} is accomplished by minimizing $\lambda_1
\, \lambda_2$.  Using \eqref{yukawa} we find the bounds on the
proto-Yukawa couplings $\lambda_i\geq (m_t/v)$ (with the lower bound reached as $\lambda_j \rightarrow \infty$)  or $\lambda_1\lambda_2 \geq 2(m_t/v)^2$. Setting
$c=\hat c'=1$ to numerically estimate the strength of the bound we
obtain $g_1^{\prime2}+g_1^2>0.99$.  The bound is a constraint on the
couplings at the Goldstone decay constant  scale $F$ of the \LLH\ theory. 
The constraint at the scale $F$ restricts the gauge coupling parameter
space available for the lower scale matching.\footnote{We will neglect the small effect of running when considering the bounds in what follows.} At approximately the EW scale ($v \approx 246 \, {\rm GeV}$) the couplings in the
\LLH\ model reduce to the SM gauge couplings as
\begin{equation}
g_{SU(2)} = \frac{g_1g_2} {\sqrt{g_1^2+g_2^2}}, \quad g_{Y} =\frac{g'_1g'_2} {\sqrt{g_1^{\prime2}+g_2^{\prime2}}}. 
\end{equation}
These relations, and the measured values of $g_{SU(2)} \, $, $g_{Y}$ ensure that if $g_i,(g_i') \rightarrow  4 \, \pi$ then $g_j, (g_j') \rightarrow g_{SU(2)}, (g_{Y})$ for the SM to be obtained in the low energy limit.\footnote{If the strong coupling limit $g_i,(g_i') \rightarrow  4 \, \pi$ is reached the ability to conclude anything in perturbation theory (including the bound) is removed. We are considering the case that the the gauge coupling is not driven into the truly strong coupling regime.}

Generically the bound will be stronger than $g_1^{\prime2}+g_1^2>0.99$
as the couplings $\lambda_1,\lambda_2$ need not take on the values
that give the minimal bound (in fact the minimal bound can only be obtained by fine tuning
these couplings). As one moves away from
the minimal case $\lambda_1\lambda_2 = 2 \, (m_t/v)^2$, the bound
grows almost {\it quadratically}, forcing a subset of the gauge couplings to
be large.  For example, for $\lambda_1\lambda_2 = 4 \, (m_t/v)^2$
the bound is given by  $g_1^{\prime2}+g_1^2>2.8$. In addition the bounds depend on the precise 
values of $c,\hat c'$ and if $c < 1$ then the bounds are stronger.

Consider the following example of the consequences of the top misalignment bound. One of the main limitations of the \LLH\ model stems from the amount
of custodial symmetry violation present in the model. The largest effect of custodial symmetry violation has a dependence on the gauge couplings of the form \cite{Han:2003wu}
\begin{eqnarray}
\rho =  1 + \frac{v^2}{F^2} \, \frac{5}{4} \left(\frac{g_1^{\prime2}-g_2^{\prime2}}{g_1^{\prime2}+g_2^{\prime2}} \right)^2 - \frac{v'^2}{v^2},
\end{eqnarray}
where $v'$ is the triplet scalar vev.\footnote{We neglect the weak
  dependence of $v'$ on the gauge couplings.} Custodial symmetry is
minimized when there is a degeneracy of
the gauge couplings of the form $g_1^{\prime2} \sim
g_2^{\prime2}$. Consider this phenomenologically favored region of
parameter space where
$g_2' = g_1'(1+ a)$ with $a \ll 1$. Then the bound
translates into the following new constraint (using
$g_{Y} \approx 0.35$)
\begin{eqnarray}
\frac{g_{1}^2}{g_1^{\prime2}} \geq  3.1+ 4.1 \, a + \mathcal{O}(a^2)).
\end{eqnarray}

The top quark induced vacuum alignment bounds do remove some of the
\LLH\ parameter space with minimal custodial symmetry violation.  In
the remaining region of favored parameter space, due to the bound, hierarchies must
exist among the \LLH\ gauge couplings for $\Sigma_{ew}$ to be the vacuum of the model. As the bound grows almost
quadratically as one moves away from the minimal value of $\lambda_1
\, \lambda_2$, we see that the top misalignment bounds 
select for a gauge coupling hierarchy of the form $g_{1} >
g_1^{\prime} \sim g_2^{\prime} > g_2$ in this favored region of
parameter space.

This is an interesting dynamical mechanism, the minimization of the
model's energy in selecting the correct vacuum alignment  ($\Sigma_{ew}$) 
requires that a subset of the gauge couplings are large. This idea of a vacuum alignment induced
gauge coupling hierarchy can have significant phenomenological
consequences as we discuss in the next section.

The top misalignment bound is of marginal interest in the
\LLH\ model alone as this model is already so phenomenologically
disfavored by EWPD.  Further, the bounds can be
satisfied and the SM couplings obtained by choosing the \LLH\ gauge
and proto-Yukawa couplings judiciously, although this does
increase the amount of tuning in the model. As the \LLH\ model is the
template on which many LH models are based, it is of interest to
investigate these considerations in more phenomenologically viable LH
models.

\section{Top vacuum misalignment for $SU(2) \times SU(2) \times U(1)$ }
\label{singleU1}
The $SU(2) \times SU(2) \times U(1)$ LH model is an interesting LH variant. The constraints of EWPD are significantly relaxed in this model compared to the \LLH\
model \cite{Han:2003wu}. By only gauging a single $U(1)$, the heavy $U(1)$ gauge boson
of the \LLH\ model is eliminated. This improves the viability of the
model as the heavy $U(1)$ gauge boson
leads to the $\mathcal{O}(v^2/F^2)$ custodial symmetry violation, and
is in fact not very heavy. Its nonobservation at the Tevatron
\cite{Han:2003wu} requires a large Goldstone decay constant scale for the \LLH\ model and
more fine tuning in the Higgs mass. By only gauging one $U(1)$, the agreement of the model with EWPD is improved at the cost of not removing all one loop quadratic divergences in the Higgs mass. However the remaining quadratic divergence due to the $U(1)$ charge does not lead to 
significant fine tuning in the Higgs mass for a cut off of $\Lambda \sim 10 \, {\rm TeV}$. In particular, this residual amount of tuning is reduced for particular choice of the gauge couplings ($g_{1} \gg g_{2}$) in the model.
Let us examine this choice of couplings in the light of the top misalignment bound.

The $SU(2) \times SU(2) \times U(1)$ LH model is substantially the
same as \LLH\ except that the generator of the gauged $U(1)$ is given
by
\begin{equation}
Y_1=\frac1{2}\diag(1,1,0,-1,-1)
\end{equation}

As ${\cal L}_{\text{top}}$ is identical for the $SU(2) \times SU(2)
\times U(1)$ model, Equations (4-12) are unchanged. The gauge boson
mass spectrum is however different and the corresponding contribution
to the difference of the vacuum energies is
\begin{equation}
V_{w}(\Sigma_{ew})-V_w(\Sigma_{B})=-\frac{3}{16}c \, F^4\left[3 \, g_Y^2 + g_1^2\right]
\end{equation}
leading to the bound (for  $\Lambda\gg \lambda_{1,2}F$) :
\begin{equation}
\label{eq:bound2}
3 \, g_Y^2+g_{1}^2>\frac{2 \, N_c}{3\pi^2c} \, \lambda_1^2 \, \lambda_2^2
\left[2 \, \ln\left(\frac{4 \, \pi \, m_t}{\lambda_1 \, \lambda_2 \, v}\right)+\frac{\hat
    c'}2\right]
\end{equation}
$F$ as low as a TeV is consistent with EWPD in
this model \cite{Han:2003wu}. Such a low $F$ (and a correspondingly low $\Lambda$) requires $g_{1} \gg g_{2}$ which was theoretically unsatisfying as this hierarchy of gauge couplings had no explanation in the model. The top vacuum misalignment
bound supplies a condition that translates into this hierarchy of gauge couplings.

Consider the alignment bound and let $\lambda_1\lambda_2  = 2 \,  x  \, (m_t/v)^2$ where $ x \geq 1$. The bound on the $g_{1}$ coupling is 
\begin{equation}
g_1^2 > 0.99  \, x^2 \left( 1 - 0.4 \, \log{x}\right) - 3  \, g_Y^2.
\end{equation}
No constraint requires that $g_{2}$ must be as large as $g_{1}$. Further, to reduce to the SM's measured value of $g_{SU(2)}$, as $g_{1} \rightarrow 4 \, \pi$ one must have $g_2 \rightarrow g_{SU(2)}$. Thus, minimizing the energy of the system (with the constraint that the SM is obtained) and 
reducing the tuning on the proto-Yukawa
couplings drives the hierarchy in the gauge coupling $g_{1} \gg
g_{2}$. This in turn drives the allowed $F$ down to
a $\rm TeV$ in this model, which in turn reduces the remaining tuning
required for the Higgs mass!

Generic signals of LH models at LHC include the observation of the heavy top partner and 
the new heavy gauge bosons and have been extensively studied in the literature \cite{Han:2003wu}. 
Other LH models can have very similar phenomenology (accessible at LHC) to the $SU(2) \times SU(2) \times U(1)$ model. It has also been noted that to discriminate between LH models at LHC and to distinguish them from  supersymmetry and extra dimension scenarios can be challenging, although strategies exist in the literature  \cite{Belyaev:2008pk}.

We note that when the scale $F \sim {\rm TeV}$ in this model, one expects the
properties of the Higgs to deviate from its properties in the SM
significantly. This is another source of experimental information to aid in the discrimination between models.  Integrating out all of the details of this LH model to
study physics around the EW scale, one obtains dimension six
operators that are suppressed by $v^2/F^2$ modifying the
properties of the pseudo Goldstone Higgs. The operators that are induced at
tree level can have interesting
phenomenological effects\footnote{Such low energy effects on 
Higgs phenomenology can also be the source of  the baryon-antibaryon asymmetry of
the universe at the modified electroweak phase transition in the
Pseudo-Goldstone Baryogenesis scenario \cite{Delaunay:2007wb}.},
such as enhancing the $g \, g \rightarrow h \, h$
signal at LHC allowing one to measure the Higgs self coupling \cite{Grinstein:2007iv}.
Operators of this form can also significantly effect the decay width of the low mass Higgs \cite{Mantry:2007ar} ($m_h < 140 \, {\rm GeV}$) suppressing or enhancing the low mass Higgs discovery signal $gg \rightarrow h \rightarrow \gamma \, \gamma$. Other dimension six operators that this model will induce (suppressed by loop factors) can also effect the production  
mechanism of the low mass Higgs at LHC and ILC \cite{Gounaris:1998ni}.  The Wilson coefficients of all of these operators are restricted by the constraint that the top misalignment bound places on the couplings of the model.

\section{Top vacuum alignment bounds and T Parity.}
Another approach to improving the viability of LH constructions is to impose T parity \cite{Cheng:2003ju}
under which all the new heavy gauge bosons and the scalar triplet are odd. This forbids the 
tree level contributions of these new states to EWPD observables. 

In order to study the effect of the top quark on vacuum alignment in
LH models with $T$-parity, a transformation law for the vacuum
orientation matrix $\Sigma$ is needed. Rather than attempting a
general argument we content ourselves with a specific example. We
consider a model with the same global and gauge symmetries as the
\LLH. To elucidate the transformation properties of $\Sigma$ under
$T$-parity we consider a UV completion consisting of a theory with
five spinors in a real representation $R$ of a techni-strong gauge
interaction such that the symmetric part of  $R\times R$ contains a
singlet. Collect the spinors into two doublets, $\psi_\pm$, with
$\psi_+$ ($\psi_-$) a doublet of $SU(2)_1$ ($SU(2)_2$), and a singlet
$\psi_0$. Since the action of $T$ parity exchanges $SU(2)_1\times
U(1)_1$ 
with $SU(2)_2\times U(1)_2$, the spinors transform, up to
trivial unitary redefinitions, by 
\begin{equation}
\psi_\pm\to \psi^c_\mp,\qquad\psi_0\to\psi_0^c,
\end{equation} 
where the  superscript ``c'' denotes charge conjugation. This allows a
complete characterization of the transformation of the condensate
$\vev{\psi\psi}$, but for our purposes it is only necessary to note that if
the upper left and lower right $2\times2$ blocks of $\Sigma$ are
denoted by $A_\pm$, then under $T$-parity $A_\pm\to A_\mp^\dagger$.

The leading  (quadratically divergent) term in the top-induced vacuum energy
in \eqref{eq:Vpsi2} depends on $\Sigma$ through a positive definite
function of $A^\dagger_-A_-$. Both $\Sigma_{ew}$ and $\Sigma_B$ alignments  in the
\LLH\ case are obtained precisely by setting $A_-=0$. When the
top-quark sector is extended to insist on $T$-parity symmetry, the
resulting vacuum energy is symmetric under the exchange  $A_\pm\to
A_\mp$. 
The vacuum energy is a sum of two positive definite functions
one of $A_+^\dagger A_+$ and one of $A^\dagger_-A_-$. The minimum
energy is obtained by setting $A_+=A_-=0$. The vacuum alignment with this property is up to gauge rotations  the  $\Sigma_{ew}$ vacuum alignment. 

There is no top induced misalignment regardless of the
strength of the interactions in $G_w$. In fact, one could consider
models in which vacuum alignment is completely driven by the
top sector. A curious example has $G_w$ replaced by a single
$SU(2)\times U(1)$. The collective symmetry argument applies exactly
in the gauge sector so the higgs potential arises only from the
top-quark sector induced Coleman-Weinberg potential. However, in this
model $SU(2)\times U(1)$ is completely broken at the scale
$F$. Alternatively one can use for the gauged group the diagonal sum
of the generators of the \LLH\ model. This remains unbroken, but has
no collective symmetry to prevent radiative contributions to the higgs
mass quadratic in $F$.

\section{Cosmological considerations}
We have derived our bound insisting on  absolute stability of the
 $\Sigma_{ew}$ vacuum as it is likely that as the universe cools
 it picks the stable vacuum.  In a cosmological setting one may
 relax this condition by considering metastability \cite{Coleman:1977py}.

Consider the case that the bound \eqref{eq:bound1} is violated and the $\Sigma_{ew}$ vacuum
is selected as the universe cools down in the \LLH \,  model.
If this occurs the $\Sigma_{ew}$ vacuum solution is not the absolute minimum vacuum solution and is metastable. A reliable computation of the lifetime of the
metastable vacuum requires understanding  the shape of the potential,
which depends on many field variables. This is beyond the scope of
this work. A rather crude estimate is obtained as follows. We take the
height of the potential barrier to be  $F^4$ and the distance
on field space between the vacua to be $F$. Then the condition that
the metastable vacuum does not decay within the age of the universe,
$t_0$, is 
\begin{equation}
\frac{(t_0F)^4}{4\pi^2}\left(\frac{k}{\epsilon^3}\right)^2e^{-k/\epsilon^3}
\lesssim1 
\end{equation}
where $k=(4\pi)^2(16/3)^3$ and $\epsilon$ is the difference between
the right hand side and the left hand side of
\eqref{eq:bound1}. This condition is satisfied
for a wide range of couplings that violate our bound. However, for our
bound to be violated this way requires the $\Sigma_{ew}$ vacuum to be
selected over the $\Sigma_B$ vacuum as the universe  cools
down. Determining whether this is the case requires knowledge of the
UV completion and thermal evolution of the theory.

By the same token, consider the case that  the bound \eqref{eq:bound1} is satisfied and yet the universe cools down into the $\Sigma_{B}$ vacuum. Then the $\Sigma_{B}$  vacuum solution would be metastable. 
The lifetime of this metastable vacuum is determined just as above and 
hence can easily be longer than the age of the universe. This
possibility is largely insensitive to the particular parameters of the theory. 
 
To the extent that the crude estimate of the lifetime of the metastable vacua
is reliable, we conclude that it is the thermal evolution that
determines the  vacuum of the universe when metastability is considered.  

\section{Conclusions}

We have shown that the effect of the top quark on vacuum alignment can dominate the effect of the 
gauge interactions in selecting the vacuum alignment of LH theories. We have demonstrated that to ensure that the low energy limit of the LH model reduces to the standard model, one can derive bounds on the size of particular LH model's gauge couplings. We have derived such bounds for the \LLH\ and $SU(2) \times SU(2) \times U(1)$ LH models. 

In examining the consequences of these bounds, we have observed that they require that a subset of gauge couplings of the LH model must be large. Such hierarchies in gauge couplings can 
have significant phenomenological consequences; the hierarchy $g_{1} \gg g_{2}$ in the 
$SU(2) \times SU(2) \times U(1)$ LH model that is phenomenologically appealing can be justified by the top induced vacuum misalignment bound.

By considering a particular UV completion on LH models with T parity, we have argued that 
the top misalignment bound does not constrain the gauge couplings of the model.

This work can be extended in many ways.  The effect of 
top induced vacuum misalignment on many other LH models should be examined. Further, one can 
examine the bounds at higher order in the gauge and proto-Yukawa couplings to further refine the bounds and the effect of running between the scale $F$ and the scale $v$ can be incorporated. 
It would also be interesting to use the top sector to drive the vacuum alignment in model building where T parity is imposed and to investigate further the possibility of metastability.

\vspace{2cm}
\begin{center}
{{\bf{Acknowledgments}}}
\end{center}
Work supported in part by the US Department of Energy under contract DE-FG03-97ER40546.

\end{document}